    \documentclass[mathleft
    ]{an}  
    \usepackage{graphicx}  
    \usepackage{times} 
    \usepackage{graphicx}
    \usepackage{aas_macros} 
    \usepackage{amssymb}
    \usepackage{rotating}                             
    \usepackage{multirow}
    \usepackage[breaklinks,colorlinks=true,citecolor=black,linkcolor=black,urlcolor=blue,bookmarks=true]{hyperref}
    \usepackage{breakurl}  
    \usepackage{amsmath}

    \usepackage{natbib}
    \newcommand{\kms}{km\,s$^{-1}$}               
    
    \newcommand{\magb}{M$_{\emph{B}}$}
    \newcommand{\feh}{$[Fe/H]$}
    \newcommand{\mgf}{$[Mg/Fe]$}
    \newcommand{\zh}{$[Z/H]$}
    \newcommand{\afe}{$[\alpha/Fe$]}

\begin{document}

    \Pagespan{960}{}
    \Yearpublication{2009}
    \Volume{330}
     \DOI{10.1002/asna.200911272}%

    \title{Metallicity  Gradients - Mass  Dependency  in  Dwarf  Elliptical
      Galaxies}

    \author{M.    Koleva\inst{1,2,3}\fnmsep\thanks{Corresponding   author:
        \email{koleva@iac.es}\newline}
    \and   Ph.   Prugniel\inst{3}   \and   S. De  Rijcke\inst{4}   \and
           W. W. Zeilinger\inst{5}  \and   D. Michielsen\inst{4}    }
    
\titlerunning{Metalllicity Gradients in Dwarf Galaxies}
\authorrunning{M. Koleva   et.al.}   

\institute{   Instituto   de
      Astrof\'{\i}sica  de Canarias,  La Laguna,  E-38200  Tenerife, Spain
      \and  Departamento de  Astrof\'{\i}sica, Universidad  de  La Laguna,
      E-38205  La  Laguna,   Tenerife,  Spain  \and  Universit\'e  Lyon~1,
      Villeurbanne, F-69622, France; CRAL, Observatoire de Lyon, St. Genis
      Laval,   F-69561,  France;   CNRS,  UMR   5574   \and  Sterrenkundig
      Observatorium, Ghent  University, Krijgslaan 281,  S9, B-9000 Ghent,
      Belgium  \and  Institut  f\"{u}r Astronomie,  Universit\"{a}t  Wien,
      T\"{u}rkenschanzstra{\ss}e 17, A-1180 Wien, Austria }

    \received{2009 Jul 14} \accepted{2009 Sep 14} \publonline{2009 Oct 20}


    \abstract{%
      The formation and evolution of galaxies is imprinted on their
      stellar population radial gradients.  Two recent articles
      present conflicting results concerning the mass dependence of
      the metallicity gradients for early-type dwarf galaxies. On one
      side, Spolaor et al.  show a tight positive correlation between
      the total metallicity \zh{} and the mass.  On the other side, in a
      distinct sample\thanks{Based on observations made with ESO telescopes at La Silla Paranal observatory under program ID076.B-0196}, we do not find any trend involving \feh{} 
      (Koleva et al.). In order to investigate the origin of the
      discrepancy, we examine various factors that may affect the
      determination of the gradients: namely the sky subtraction and
      the signal-to-noise ratio. We conclude that our detection of
      gradients are well above the possible analysis biases. Then, we
      measured the \mgf{} relative abundance profile and found
      moderate gradients.  The derived \zh{} gradients scatter around
      -0.4\,dex/$r_e$.  The two samples contain the same types of
      objects and the reason of the disagreement is still not
      understood.  }  \maketitle

\sloppy
    \section{Introduction}
    The stellar populations of galaxies  hold a fossil record of their
    formation and  evolution.  Their  study may allow  to discriminate
    between  the  various  formation  scenarios  predicting  different
    spatial  gradients and different  relations between  the gradients
    and the mass of the galaxy.

    In  the  classical  monolithic  scenario,  galaxies  formed  in  a
    dissipative collapse \citep{lar1974,ay1987}, where stars remain on
    their orbits and do not mix.  The gas, enriched in metals from the
    evolved  stars,   flows  into  the   centre,  generating  negative
    metallicity  gradients (\emph{i.e.}  higher  metal content  in the
    centre  than in the  outskirts).  The  chemo-dynamical simulations
    \citeauthor{mat1987}  \citep[1987,  see also][]{kaw2003}  predict,
    that due to  their small potential, dwarf galaxies,  are unable to
    retain  metals and thus  their gradients  are shallower  or almost
    inexistent.   Therefore  relations  between   the  mass   and  the
    metallicity or the metallicity gradient are expected.

    A   competing  (or  complementary)   formation  scenario   is  the
    hierarchical  clustering \citep[e.g.][]{col1994}, where  small dark
    matter halos  merge to produce bigg\-er ga\-la\-xi\-es.  It may be
    thought that  these violent events mix the  stellar population and
    erase the  gradients \citep[e.g.][]{whi1980}, but  some simulations
    \citep[e.g.][]{dim2009}, at variance, show a gradient preservation,
    due to the violent relaxation which does not mix the orbits.

    The   interpretation   of  the   metallicity   gradients  is   not
    straightforward.   Nevertheless,   gradients   observation   could
    constrain model's  predictions. Several studies on  big samples of
    giant   elliptical   galaxies    have   been   already   performed
    \citep[e.g.][]{psb2006}.   However,   due   to  their   low-surface
    brightness,  the  dwarf  galaxies,  are  still  very  much  'terra
    incognita'.
    
\begin{figure}[ht!]
      \includegraphics{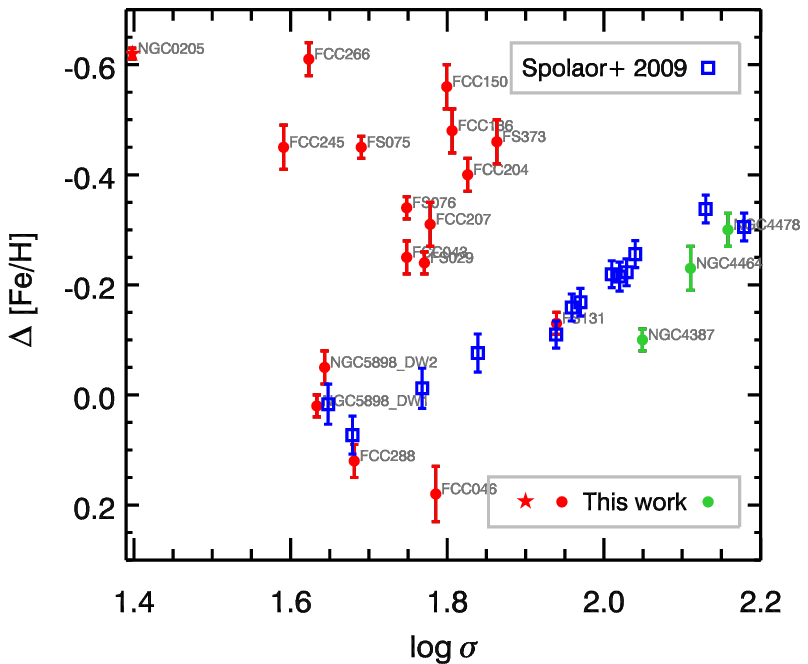} \includegraphics{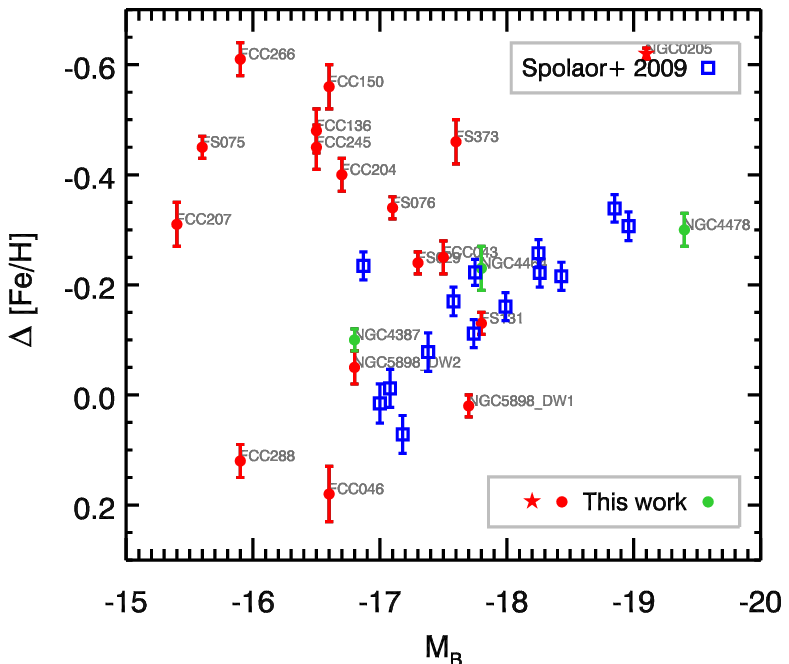}
      \caption[]{   Metallicity  gradients  \emph{vs.}   $\sigma$  and
        \magb{}  (top and  bottom respectively)  of the  2  samples of
        dwarf  elliptical  galaxies,   one  from  \citet[][blue,  open
          squares]{spolaor2009}  and   other  is  this   work  (filled
        symbols).  Green  circles are galaxies  from \citet{vaku}, red
        circles   from   \citet{kol2009}    and   a   red   star   for
        NGC\,205\citep{sp2002}.   The  names of  our  galaxies are  also
        plotted.}
      \label{fig:gradients}
    \end{figure}

    Recently we published  a paper on the stellar  content of 16 dwarf
    galaxies in the Fornax  cluster and in groups \citep{kol2009}. One
    of  our main  conclusions was  that most  of our  galaxies  hold a
    strong negative metallicity  gradient.  The star formation history
    analyses  revealed  that  these   gradients  are  built  at  early
    epochs. We did not find  any correlation between our gradients and
    the  velocity  dispersion  (proxy  of  the  mass)  in  our  sample
    (Fig.\ref{fig:gradients}).  However, this  result is challenged by
    a  recent letter from  \citet{spolaor2009}, using  a sample  of 14
    dwarfs in the Virgo and  Fornax clusters observed with GMOS at the
    Gemini  South telescope in  medium resolution,  long-slit, optical
    spectroscopy.

    To analyse  their data, \citet{spolaor2009} binned  them along the
    slit  direction  to achieve  a  minimum  signal-to-noise  of 30  @
    5200\,\AA.   They measured  Lick indices  \citep{wor1994,wo1997} and
    derived  ages  and  chemical  compositions  by  comparing  to  the
    \citet{tho2004}  single  stellar  population (SSP)  models,  using
    their  own method \citep{pro2002}.   They estimated  the gradients
    performing a  linear least-squares  fit to the  radial metallicity
    profile weighted by their errors.  The fit was done in logarithmic
    space  of the  radius, $\Delta$\zh/$\Delta\log(r/r_e)$,  where $r$
    varies from 1 arcsec to one effective radius ($r_e$).

    There  is no  galaxy in  common between  the two  samples  and the
    causes of the disagreement may  be multiple: choice of the sample,
    data reduction  or analysis issue.   The primary purpose  of the
    present  paper is  to explore  these possibilities.   In  order to
    directly compare our results  with \citet{spolaor2009} we will determine
    our metallicity gradients in an identical way.

    In Sect.2  we present the  sample, observations and  analysis.  In
    Sect.\ref{section:snr}    we    investigate     the    systematics
    (signal-to-noise ratio, bad  sky-subtraction) which may affect the
    derived  metallicity gradients.  As  \citeauthor{spolaor2009} used
    the total metallicity \zh we  also measured \mgf{} and discuss the
    differences   between    $\Delta$\feh{}   and   $\Delta$\zh{}   in
    Sect.\ref{subsec:feh}.

    \section{Data and analysis}
    For the  purpose of the  present paper we collected  20 early-type
    galaxies (Table \ref{table:1}) with velocity dispersions less than
    150\,\kms{} from three different data  sets (VLT, OHP and WHT). We
    analysed them with the same method.
     
    \begin{table*}
    \caption{Properties  of our galaxies sample.  In column  1 we
      list  the galaxies'  names, in  columns 2  and 3  the equatorial
      coordinates,  in column 4  the \emph{B}-band  apparent magnitude
      m$_{\emph{B}}$,  in column  5 the  Schlegel  galactic extinction
      \citep[][\textsc{abg},\textsc{pleinpot}]{sch1998},  in  column 6
      the  radial velocities (measured  with ULySS),  in column  7 the
      k-correction (from  \textsc{cocor}, \textsc{pleinpot}, using the
      listed cz  values), in column 8 the  absolute magnitudes \magb{}
      in \emph{B}-band (computed as  described in the text), in column
      9  the effective  radius derived  from the  \emph{R}-band images
      \citep{der05},  in column  10  the luminosity-weighted  velocity
      dispersion \citep{der05}, in column 11 the metallicity gradients
      (derived as described in the text). }
    \label{table:1}
    \begin{tabular}{lcccrrrrrrr}
    \hline 
  \multicolumn{1}{l}{Object}         & \multicolumn{1}{c}{$\alpha_{J2000}$} & \multicolumn{1}{c}{$\delta_{J2000}$}  
& \multicolumn{1}{c}{m$_{\emph{B}}$} & \multicolumn{1}{c}{A$_{\emph{B}}$}   & \multicolumn{1}{c}{$cz$}        
& \multicolumn{1}{c}{K}              & \multicolumn{1}{c}{\magb}            & \multicolumn{1}{c}{r$_e$}          
& \multicolumn{1}{c}{$\sigma$}       & \multicolumn{1}{c}{$\Delta$[Fe/H]} \\
  \multicolumn{5}{c}{ }         
& \multicolumn{1}{c}{[\kms]}        
& \multicolumn{2}{c}{ }              
& \multicolumn{1}{c}{[$''$]}          
& \multicolumn{1}{c}{[\kms]} 
& \multicolumn{1}{c}{[dex]} \\
\hline
NGC\,205      & 00 40 22.1 & +41 41 07 &  8.89$^a$ & 0.37  &  -234 & +0.00 & -19.1 & 150.0$^b$ &  25$^c$  & -0.62 $\pm$ 0.01\\
NGC\,4387      & 12 25 41.7 & +12 48 38 & 12.97$^a$ & 0.14  &  +586 & -0.01 & -16.8 &  26.0$^d$ & 112$^d$  & -0.10 $\pm$ 0.02\\
NGC\,4464      & 12 29 21.3 & +08 09 24 & 13.56$^a$ & 0.09  & +1266 & -0.02 & -17.8 &   6.0$^d$ & 129$^d$  & -0.23 $\pm$ 0.04\\
NGC\,4478      & 12 30 17.8 & +12 19 37 & 12.16$^a$ & 0.10  & +1390 & -0.02 & -19.4 &  17.0$^d$ & 144$^d$  & -0.30 $\pm$ 0.03\\
FCC\,043       & 03 26 02.2 & -32 53 40 & 13.91 & 0.06  & +1337 & -0.02 & -17.5 &  16.9 &  56  & -0.25 $\pm$ 0.03\\
FCC\,046       & 03 26 25.0 & -37 07 41 & 15.99 & 0.08  & +2252 & -0.03 & -16.6 &   6.7 &  61  & +0.18 $\pm$ 0.05\\
FCC\,136       & 03 34 29.5 & -35 32 47 & 14.81 & 0.07  & +1238 & -0.02 & -16.5 &  14.2 &  64  & -0.48 $\pm$ 0.04\\
FCC\,150       & 03 35 24.1 & -36 21 50 & 15.70 & 0.06  & +2000 & -0.03 & -16.6 &   5.7 &  63  & -0.56 $\pm$ 0.04\\
FCC\,204       & 03 38 13.6 & -33 07 38 & 14.76 & 0.04  & +1368 & -0.02 & -16.7 &  11.5 &  67  & -0.40 $\pm$ 0.03\\
FCC\,207       & 03 38 19.3 & -35 07 45 & 16.19 & 0.06  & +1429 & -0.02 & -15.4 &   8.4 &  60  & -0.31 $\pm$ 0.04\\
FCC\,245       & 03 40 33.9 & -35 01 23 & 16.00 & 0.05  & +2180 & -0.03 & -16.5 &  11.4 &  39  & -0.45 $\pm$ 0.04\\
FCC\,266       & 03 41 41.4 & -35 10 12 & 15.90 & 0.05  & +1546 & -0.02 & -15.9 &   7.1 &  42  & -0.61 $\pm$ 0.03\\
FCC\,288       & 03 43 22.6 & -33 56 25 & 15.10 & 0.03  & +1087 & -0.02 & -15.9 &   9.5 &  48  & +0.12 $\pm$ 0.03\\
FS\,029        & 13 13 56.2 & -16 16 24 & 15.70 & 0.28  & +2447 & -0.04 & -17.3 &   8.9 &  59  & -0.24 $\pm$ 0.02\\
FS\,075        & 13 15 04.1 & -16 23 40 & 16.87 & 0.31  & +1889 & -0.03 & -15.6 &   6.8 &  49  & -0.45 $\pm$ 0.02\\
FS\,076        & 13 15 05.9 & -16 20 51 & 16.10 & 0.30  & +2698 & -0.04 & -17.1 &   4.4 &  56  & -0.34 $\pm$ 0.02\\
FS\,131        & 13 16 49.0 & -16 19 42 & 15.30 & 0.35  & +2556 & -0.04 & -17.8 &   8.1 &  87  & -0.13 $\pm$ 0.02\\
FS\,373        & 10 37 22.9 & -35 21 37 & 15.60 & 0.49  & +2444 & -0.04 & -17.6 &   7.9 &  73  & -0.46 $\pm$ 0.04\\
NGC\,5898\_DW1 & 15 18 13.0 & -24 11 47 & 15.66 & 0.61  & +2467 & -0.04 & -17.7 &   8.7 &  43  & +0.02 $\pm$ 0.02\\
NGC\,5898\_DW2 & 15 18 44.7 & -24 10 51 & 16.10 & 0.66  & +1993 & -0.03 & -16.8 &   5.9 &  44  & -0.05 $\pm$ 0.03\\
\hline
    \end{tabular}
    $^a$ from HyperLeda database \citep{ps1996};$^b$ from \citet{sp2002};$^c$ from  \citet{phd}; $^d$ from \citet{vaku}
    \end{table*}

    \subsection{OHP data}
    We  used long  slit data  of NGC\,205  \citep{sp2002},  which were
    taken   with  CARELEC   spectrograph  on   1.93\,m   telescope  in
    Observatoire  de  Haute-Provence  (OHP).   The  observations  were
    carried from January  2001 to January 2003 and  are partly used in
    \citet{sp2002} to derive the  internal kinematics.  The slit width
    was set to 1.5$^{\prime\prime}$,  which results in an instrumental
    broadening       of        R\,=\,5050       (       FWHM,       or
    $\sigma_{ins}\,\approx\,25$\,\kms).   The wavelength  range covers
    $\lambda\lambda$  4700 --  5600\,\AA.  NGC\,205  is  an emblematic
    galaxy for the dwarf elliptical class.

    \subsection{VLT data}

    We  obtained long slit  spectra of  16 dwarfs  elliptical galaxies
    (dE) with FORS1,2 mounted on the VLT.  These galaxies are situated
    in the  Fornax cluster and the NGC\,5044,  NGC\,5898 and NGC\,3258
    groups, \emph{i.e.}  in different environments.  This sample covers
    a diversity  of dwarfs: nucleated, non-nucleated,  with or without
    spiral substructures, with or without gas \citep[table1]{kol2009}.
    These  dwarfs have \magb{}  between -17.8  and -15.4,  while their
    velocity  dispersion\footnote{luminosity  weighted  mean  velocity
      dispersion from \citealt{der05}} ranges from 39.5 to 87.0\,\kms.
    In other words they are similar to the dEs prototype - NGC\,205.

    The   group   galaxies  were   observed   with   FORS2  with   the
    GRIS\-1200g+96     grism,     resulting    in     $\lambda\lambda$
    3300-6200\,\AA{} and $\sigma_{ins}$\,=\,74\,\kms.  The Fornax data
    were observed with the GRIS  600B+22 grism on FORS1, which results
    in          $\lambda\lambda$          4335-5640\,\AA{}         and
    $\sigma_{ins}$\,=\,64\,\kms at 5200\,\AA.

    \subsection{VAKU sample}

    We also  used three  of the 6  Virgo cluster galaxies  analysed in
    \citet{vaku}.  These three objects  are situated in the transition
    region    between   normal    and   dwarf    elliptical   galaxies
    ($\sigma\,\approx\,$120\,\kms  and \magb\,$\sim$\,-18).   The long
    slit observations  were made  with the William  Herschel Telescope
    (WHT)    in    La    Palma.     The   wavelength    coverage    is
    $\lambda\lambda\,\approx\,4000-5000\,\AA$,  and  the  instrumental
    resolution is 2.4\,\AA{}, or R$\,\approx\,$2000.

    \subsection{Analysis}

    The SSP-equivalent  profiles (see Fig. 1,  2 of \citealt{kol2009})
    were obtained by binning the  2D spectrum in the spatial direction
    to  achieve a  signal-to-noise ratio  of 20  and by  analysing the
    resulting               1D               spectra              with
    ULySS\footnote{\url{http://ulyss.univ-lyon1.fr}}
      \citep[Universit\'e     de    Lyon     Spectroscopic    analysis
        Software,][]{ulyss}.

    ULySS  is designed  to fit  any linear  combination  of non-linear
    components    (weighted    by     $W$)    against    a    spectrum
    $F_{obs}(\lambda)$. A model constructed  as such, can be convolved
    with a line-of-sight  velocity distribution (LOSVD) and multiplied
    by a $n^{th}$ order Legendre polynomial, $P_n(\lambda)$.:
    \begin{align}
    F_{obs}(\lambda)    =   P_{n}(\lambda)    \times    \bigg(   &{\rm
      LOSVD}(v_{sys},\sigma,    h3,   h4)   \nonumber    \\   &\otimes
    \sum_{i=0}^{i=k}  W_i \,\,  {\rm  CMP}_i\,(a_1, a_2,  ...,\lambda)
    \bigg).
    \label{eq:main}
    \end{align}

    \noindent  The  LOSVD is  a  function  of  the systemic  velocity,
    $v_{sys}$ and  the velocity dispersion $\sigma$ and  may include a
    Gauss-Hermit   expansion  ($h3$   and   $h4$,  \citealp{marel93}).
    $\lambda$  is the  logarithm  of the  wavelength (the  logarithmic
    scale  is  required  to express  the  effect  of  the LOSVD  as  a
    convolution).  In  this particular case, the CMP  ($i\,=\,1$) is a
    SSP  constructed   with  P\'egase.HR  stellar   population  models
    \citep{phr}    using   Elodie.3.1   empirical    stellar   library
    \citep{elo1, elo3.1}.  Consequently, the parameters $a_1, a_2$ are
    the age and metallicity of the  SSP. For the $\alpha$-elements analysis we
    added   another  parameter  $a_3$   which  expresses   the  \mgf{}
    abundance.  In the latter  case the stellar population models were
    build with  a semi-empirical  library, produced by  combination of
    Elodie.3.1 and \citet{coe2005}  libraries, including the non-solar
    abundance   of  \mgf{}  \citep{php2007,kol2008}.  More  precisely,
    Elodie.3.1 was differentally corrected for \mgf{}\,=\,0.0 and 0.4,
    using  \citeauthor{coe2005}  library.   The  analysis  selects the
    optimal \mgf{} value interpolated in this range.

    Some of the galaxies formation models predict relation between the
    mass of a  galaxy and its metallicity gradient. As  a proxy to the
    mass we  used the velocity dispersion and  the absolute magnitude.
    We took  the velocity dispersion  values from the  literature (see
    Table\ref{table:1}).  The absolute  magnitudes were computed using
    $cz$, $m_{\emph{B}}$ (from Table\ref{table:1}) and Hubble constant
    H$_0$\,=\,70  \kms\,Mpc$^{-1}$.  We corrected  for \citet{sch1998}
    galactic extinction and  reddening (derived with \textsc{abg} and
    \textsc{cocor}           respectively           from           the
    \textsc{pleinpot}\footnote{\url{http://leda.univ-lyon1.fr/pleinpot/pleinpot.html}}
    package).
    
    We did not find any  obvious correlation between our gradients and
    the mass of the  galaxies.  From Fig.\,\ref{fig:gradients} one can
    notice  that if  we had  data only  from \citet{vaku}  and several
    dwarfs from \citeauthor{kol2009} (FCC\,288, DW\,1, DW\,2, FS\,131)
    then we  could have found  a mass-gradient relation. We  note also
    that,  some of our  flattest galaxies  ($\epsilon\,>\,0.3$; DW\,1,
    DW\,2, FCC\,046,  FCC\,288, FS\,029, FS\,131) follow  a trend with
    \magb{}  and  $\sigma$.  However,  FCC\,204,  which  is also  flat
    ($\epsilon\,=\,0.61$),  is  an  outlier.   We cannot  exclude  the
    possibility that the contrasting results are due to the small size
    of the sample or to the actual selection critreria.

    \section{Validation}

    \subsection{Noise and sky subtraction}
    \label{section:snr}
    The two main features of the data reduction and analysis which can
    affect the  derived metallicity gradients  are a possible  bad sky
    subtraction and the decreasing  signal-to-noise from the centre to
    the  outskirts. Our  measurements are  almost insensitive  to flux
    calibration problems (as the Lick indices).  The shape of the flux
    is modeled by a  multiplicative polynomial, during the fitting and
    cannot be  a source of  biases.  

    To investigate  how the  signal-to-noise and imperfections  of the
    sky subtraction can affect the measured metallicities, we produced
    a  1D  model  of  the  centre  of  FS\,373,  using  the  P\'egase.HR
    population synthesis code.  The  choice of this particular galaxy,
    was driven  by the  its strong metallicity  gradient (-0.46\,dex).
    To construct  the model, we  use the SSP-equivalent  parameters as
    derived  from  our  analysis, \emph{i.e.}   age\,=\,1.55\,Gyr  and
    \feh\,=\,0.0\,dex.   We  broadened   the  model  to  the  velocity
    dispersion (instrumental plus physical) of the observed spectrum.

\begin{figure}[ht!]
      \includegraphics{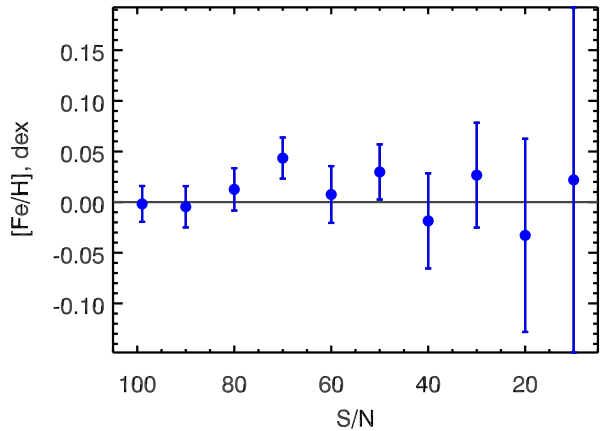}
      \caption[]{ Dependence of the ULySS metallicity determination on
        the signal-to-noise of  the data.  With grey line  we plot the
        input  metallicity and  with  blue filled  circles the  output
        values of \feh.  }
      \label{fig:feh_snr}
    \end{figure}

    To  investigate the  metallicity dependence  on the  S/N  we added
    Gaussian noise to the model using \textsc{snr} and \textsc{nsimul}
    keywords  in  the  call  of  ULySS.   The  results  are  shown  in
    Fig.\,\ref{fig:feh_snr}.   The errors become  larger when  the S/N
    decreases, an expected behaviour.   However, we do not observe any
    important systematics  in the measured metallicity  values down to
    S/N\,=\,10.

\begin{figure}[ht!]
      \includegraphics{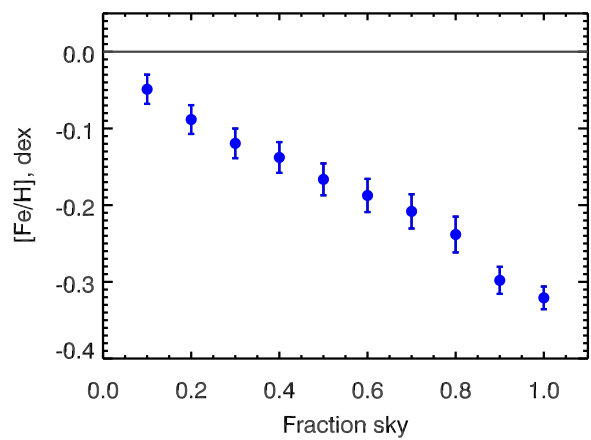}
      \caption[]{Dependence  of  the  metallicity  determination  with
        ULySS on the under-subtracted  sky in fraction.  The grey line
        marks the input metallicity of  the model, while the blue dots
        are  the  derived  metallicity.  In  abscisse,  fraction  =  1
        corresponds to the non subtraction  of the sky, and fraction =
        0 to the exact subtraction.  }
      \label{fig:feh_sky}
    \end{figure}

    The sky subtraction, especially for diffuse objects like the dwarf
    galaxies, is a particularly  critical point.  To obtain a negative
    metallicity gradient,  one needs to  systematically under-subtract
    the  background.  To  explore this  effect,  we took  an UVES  sky
    spectrum  \citep{han2003}  and broadened  it  to the  instrumental
    resolution  of  FORS2, 74\,\kms,  and  scaled  it  to the  typical
    brightness of the  dark sky.  Taking into account  the redshift of
    the  galaxy  we added  the  two  spectra  (sky plus  galaxy).   We
    simulated an under-subtraction of the sky from fully subtracted to
    un-subtracted.       The       results      are      shown      in
    Fig.\,\ref{fig:feh_sky}. To decrease the gradients by 0.2\,dex, we
    need to  under-subtract 50\,\%  of the sky.  For FS\,373,  where the
    gradient is -0.46\,dex, we would need to under-estimate the sky by
    more than  100\,\%!  We  could possibly make  an error on  the sky
    subtraction of 10\,\% in improbable  cases, but certainly not as a
    systematic under-subtraction.   Our strong gradients  cannot be an
    artifact of  a bad sky  subtraction.  At least  not in the  way we
    modeled it here.

    \subsection{Iron against metallicity abundance} \label{subsec:feh}
    
    The metallicity  of the stellar  population models are  labeled in
    \feh{} (because the stellar libraries are indexed in this manner).
    Therefore, although the stellar  evolution depends on the detailed
    abundance  \citep{sal2000},  the  full-spectrum  fitting  measures
    \feh{} rather than \zh{} (the differences in the stellar evolution
    mostly translates as biases on the age).

    \citeauthor{spolaor2009}  using \citealt{tho2004}  models, derived
    their {\zh} as a sum of Fe-like plus Mg-like elements (\emph{C, N,
      O,  Mg, Na,  Si}).   Consequently, if  the  populations are  not
    scaled-solar, the results may not be consistent with ours.
    
\begin{figure}[ht!]
      \includegraphics{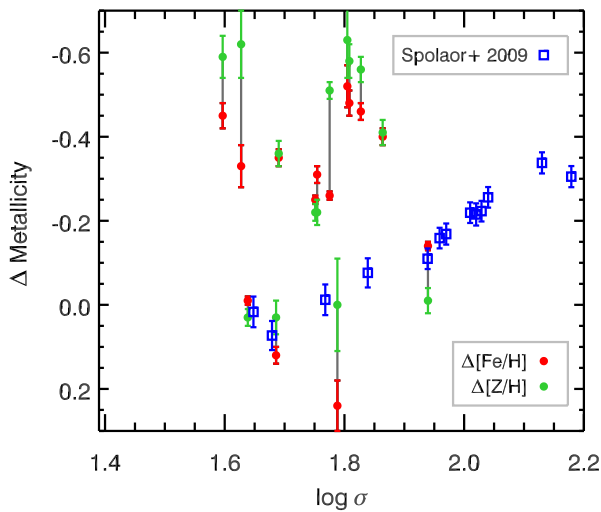}
      \caption[]{  Relation between  \feh{} (red  filled  circles) and
        \zh{}  (green filled  circles)  gradients for  the FORS  dwarf
        ellipticals. The gradients of each galaxies are connected with
        a line. The sample of \citeauthor{spolaor2009} is plotted with
        blue open squares.  }
      \label{fig:gradients1}
    \end{figure}

    To  make  the  comparison  rigorous,  we  also  derived  \zh{}  by
    measuring  \mgf.   We  explore  the possibility  that  the  \mgf{}
    gradients compensate \feh{} to  result in flat \zh{} profiles.  To
    test  the  difference  in  the  measured quantities  we  used  our
    $\alpha$-enhanced models \citep{php2007,  kol2008} as explained in
    Sect.\,2.    We  found  in   general  negative   \mgf{}  gradients
    ($\overline{\Delta[Mg/Fe]}\,=\,-0.10\pm0.09$\,dex  ).   Only  four
    galaxies (DW\,1, FS\,76, FS\,131,  FCC\,043) have flat or positive
    \mgf{}  gradient.    We  computed  the   total  metallicity  using
    \zh\,=\,\feh\,+\,0.98  \mgf{} \citep{tho2004}.  We  find that  the
    total  metallicity  gradients  (see Fig.\ref{fig:gradients1})  are
    \emph{stronger} than  the \feh{} gradients, and  the difference in
    the metallicity definitions  cannot explain the difference between
    the two data sets.

    \section{Discussions}
        
    We investigated the  discrepancy in the metallicity gradients-mass
    relation  between  two recent  works,  namely \citet{spolaor2009} and
    \citet{kol2009}. While the former  work shows a very strong and
    tight relation between the  total metallicity gradient in low-mass
    galaxies and their central  velocity dispersion or \magb, the latter
    do not find a hint of such a relation.

    As first guess  for this discrepancy we pointed  the small size of
    the samples and different  selection criteria.  Such a scenario is
    possible,    since    some     of    our    galaxies    fall    on
    \citeauthor{spolaor2009}   relation.   This   hypothesis  may   be
    supported   by   the   fact    that   the   Fornax   galaxies   of
    \citet{spolaor2009} are in  general flat ($\epsilon\,>\,0.3$), and
    we  have shown  that the  gradients in  such galaxies  are usually
    shallower.  However, in their Virgo sample all but one galaxy have
    $\epsilon\,<\,0.3$\footnote{ We  used the ellipticity measurements
      from      the      HyperLeda      database      \citep{pat2003},
      \url{http://leda.univ-lyon1.fr}}.  Hence,  this  explanation  is
    ruled out and flattening cannot be a reason for our disagreement.

    Another possible explanation would  have been that while the dwarf
    galaxies exhibit a strong  negative $\Delta$\feh, they possess the
    same \afe{}  gradient but with  a positive sign.   Thus, measuring
    the  total metallicity  as a  combination of  \feh{} and  \afe, we
    would  have a  total  $\Delta$\zh\,=\,0.  We  measured the  \mgf{}
    gradients  in our sample  using unpublished  semi-empirical models
    \citep{php2007}.   These models are  preliminary but  the relative
    measurements  of \mgf{}  are certainly  reliable. Opposite  to the
    expectations  we   found  that  (most  of)   our  galaxies possess
    \emph{negative}   \mgf{}  gradients.    Consequently,   the  total
    metallicity   gradients  are   \emph{stronger}  than   the  \feh{}
    gradients.

    Negative iron and  \mgf{} gradients will imply that,  in the early
    epochs,  when  the  feedback  is  a  strong  regulator,  the  star
    formation   starts  in   the  centre,   where  the   gas   is  the
    densest. After the gas is exhausted, the star formation stops first
    in the centre and continues  longer at larger radius. This process
    forming a  negative \mgf{} gradient.  To form  the negative \feh{}
    gradient  we need  that cool,  enriched gas  flows again  into the
    centre,  triggering  star   formation.   In  other  words,  having
    repeating  episodes  of  star  formation.   This  scenario  is  in
    agreement with  recent SPH simulations  on the formation  of dwarf
    galaxies \citep[fig.10]{val2008}

    We also investigated  the effect of a bad  sky-subtraction and low
    signal-to-noise  on our metallicity  measurements. We  showed that
    our metallicity measurements are not biased down to S/N\,=\,10 and
    that to have zero  metallicity gradients we need to under-subtract
    the sky with more than  50\,\%.  We concluded that these two effects
    cannot be the cause of our strong gradients.

    As a conclusion,  according to the tests performed  in this paper,
    the  origin of the  differences between  the two  articles remains
    unknown.

    \acknowledgements We thank the organizers for the fruitful meeting. We
    thank  Patricia S\'{a}nchez-Bl\'{a}zquez  for the  useful discussions.
    MK has  been supported by  the Programa Nacional de  Astronom\'{\i}a y
    Astrof\'{\i}sica  of the  Spanish Ministry  of Science  and Innovation
    under  grant   \emph{AYA2007-67752-C03-01}  and  Bulgarian  Scientific
    Research Fund \emph{DO 02-85/2008}. We used  
    \textsc{dexter} (\url{http://dc.zah.uni-heidelberg.de/sdexter}) to
    acquire the measurements from published figures.

    \newpage

\begin{thebibliography}{31}
\expandafter\ifx\csname natexlab\endcsname\relax\def\natexlab#1{#1}\fi

\bibitem[{{Arimoto} \& {Yoshii}(1987)}]{ay1987}
{Arimoto} N., {Yoshii} Y., 1987, \aap, 173, 23

\bibitem[{{Coelho} {et~al.}(2005){Coelho}, {Barbuy}, {Mel{\'e}ndez},
  {Schiavon}, \& {Castilho}}]{coe2005}
{Coelho} P., {Barbuy} B., {Mel{\'e}ndez} J., {Schiavon} R.~P., {Castilho}
  B.~V., 2005, \aap, 443, 735

\bibitem[{{Cole} {et~al.}(1994){Cole}, {Aragon-Salamanca}, {Frenk}, {Navarro},
  \& {Zepf}}]{col1994}
{Cole} S., {Aragon-Salamanca} A., {Frenk} C.~S., {Navarro} J.~F., {Zepf} S.~E.,
  1994, \mnras, 271, 781

\bibitem[{{De Rijcke} {et~al.}(2005){De Rijcke}, {Michielsen}, {Dejonghe},
  {Zeilinger}, \& {Hau}}]{der05}
{De Rijcke} S., {Michielsen} D., {Dejonghe} H., {Zeilinger} W.~W., {Hau}
  G.~K.~T., 2005, \aap, 438, 491

\bibitem[{{di Matteo} {et~al.}(2009){di Matteo}, {Pipino}, {Lehnert}, {Combes},
  \& {Semelin}}]{dim2009}
{di Matteo} P., {Pipino} A., {Lehnert} M.~D., {Combes} F., {Semelin} B., 2009,
  \aap, 499, 427

\bibitem[{Hanuschik} (2003)]{han2003}
{Hanuschik} R.~W., 2003, \aap, 407, 1157

\bibitem[{{Kawata} \& {Gibson}(2003)}]{kaw2003}
{Kawata} D., {Gibson} B.~K., 2003, \mnras, 340, 908

\bibitem[{{Koleva}(2009){Koleva}}]{phd} {Koleva}, M., 2009, PhD thesis

\bibitem[{{Koleva} {et~al.}(2008){Koleva}, {Gupta}, {Prugniel},
  \& {Singh}}]{kol2008}
{Koleva} M., {Gupta} R., {Prugniel} P., {Singh} H., 2008, in
  Astronomical Society of the Pacific Conference Series, Vol. 390, Pathways
  Through an Eclectic Universe, {Knapen} J.~H., {Mahoney} T.~J., {Vazdekis} A.,
  eds., pp. 302

\bibitem[{{Koleva} {et~al.}(2009{\natexlab{a}}){Koleva}, {de Rijcke}, {Prugniel},
  {Zeilinger}, \& {Michielsen}}]{kol2009}
{Koleva} M., {de Rijcke} S., {Prugniel} P., {Zeilinger} W.~W., {Michielsen} D.,
  2009{\natexlab{a}}, \mnras, 396, 2133

\bibitem[{{Koleva} {et~al.}(2009{\natexlab{b}}){Koleva}, {Prugniel},
  {Bouchard}, \& {Wu}}]{ulyss}
{Koleva} M., {Prugniel} P., {Bouchard} A., {Wu} Y., 2009{\natexlab{b}}, \aap,
  501, 1269

\bibitem[{{Larson}(1974)}]{lar1974}
{Larson} R.~B., 1974, \mnras, 166, 585

\bibitem[{{Le Borgne} {et~al.}(2004){Le Borgne}, {Rocca-Volmerange},
  {Prugniel}, {Lan{\c c}on}, {Fioc}, \& {Soubiran}}]{phr}
{Le Borgne} D., {Rocca-Volmerange} B., {Prugniel} P., {Lan{\c c}on} A., {Fioc}
  M., {Soubiran} C., 2004, \aap, 425, 881

\bibitem[{{Matteucci} \& {Tornambe}(1987)}]{mat1987}
{Matteucci} F., {Tornambe} A., 1987, \aap, 185, 51

\bibitem[{{Paturel} {et~al.}(2003){Paturel}, {Petit}, {Prugniel}, {Theureau},
  {Rousseau}, {Brouty}, {Dubois}, \& {Cambr{\'e}sy}}]{pat2003}
{Paturel} G., {Petit} C., {Prugniel} P., {Theureau} G., {Rousseau} J., {Brouty}
  M., {Dubois} P., {Cambr{\'e}sy} L., 2003, \aap, 412, 45

\bibitem[{{Proctor} \& {Sansom}(2002)}]{pro2002}
{Proctor} R.~N., {Sansom} A.~E., 2002, \mnras, 333, 517

\bibitem[{{Prugniel} {et~al.}(2007{\natexlab{a}}){Prugniel}, {Koleva},
  {Ocvirk}, {Le Borgne}, \& {Soubiran}}]{php2007}
{Prugniel} P., {Koleva} M., {Ocvirk} P., {Le Borgne} D., {Soubiran} C.,
  2007{\natexlab{a}}, in IAU Symposium, Vol. 241, IAU Symposium, {Vazdekis} A.,
  {Peletier} R.~F., eds., pp. 68--72

\bibitem[{Prugniel \& Simien}(1996)]{ps1996}
{Prugniel} Ph., {Simien} F., 1996, \aap, 309, 749

\bibitem[{{Prugniel} \& {Soubiran}(2001)}]{elo1}
{Prugniel} P., {Soubiran} C., 2001, \aap, 369, 1048

\bibitem[{{Prugniel} {et~al.}(2007{\natexlab{b}}){Prugniel}, {Soubiran},
  {Koleva}, \& {Le Borgne}}]{elo3.1}
{Prugniel} P., {Soubiran} C., {Koleva} M., {Le Borgne} D., 2007{\natexlab{b}},
  arXiv:astro-ph/0703658

\bibitem[{{Salasnich} {et~al.}(2000){Salasnich}, {Girardi}, {Weiss}, \&
  {Chiosi}}]{sal2000}
{Salasnich} B., {Girardi} L., {Weiss} A., {Chiosi} C., 2000, \aap, 361, 1023

\bibitem[{{S{\'a}nchez-Bl{\'a}zquez} {et~al.}(2006){S{\'a}nchez-Bl{\'a}zquez},
  {Gorgas}, \& {Cardiel}}]{psb2006}
{S{\'a}nchez-Bl{\'a}zquez} P., {Gorgas} J., {Cardiel} N., 2006, \aap, 457, 823

\bibitem[{{Schlegel} {et~al.}(1998){Schlegel}, {Finkbeiner}, \&
  {Davis}}]{sch1998}
{Schlegel} D.~J., {Finkbeiner} D.~P., {Davis} M., 1998, \apj, 500, 525

\bibitem[{{Simien} \& {Prugniel}(2002)}]{sp2002}
{Simien} F., {Prugniel} P., 2002, \aap, 384, 371

\bibitem[{{Spolaor} {et~al.}(2009){Spolaor}, {Proctor}, {Forbes}, \&
  {Couch}}]{spolaor2009}
{Spolaor} M., {Proctor} R.~N., {Forbes} D.~A., {Couch} W.~J., 2009, \apjl, 691,
  L138

\bibitem[{{Thomas} {et~al.}(2004){Thomas}, {Maraston}, \& {Korn}}]{tho2004}
{Thomas} D., {Maraston} C., {Korn} A., 2004, \mnras, 351, L19

\bibitem[{{Valcke} {et~al.}(2008){Valcke}, {de Rijcke}, \&
  {Dejonghe}}]{val2008}
{Valcke} S., {de Rijcke} S., {Dejonghe} H., 2008, \mnras, 389, 1111

\bibitem[{{van der Marel} \& {Franx}(1993)}]{marel93}
{van der Marel} R.~P., {Franx} M., 1993, \apj, 407, 525

\bibitem[{{Vazdekis} {et~al.}(2001){Vazdekis}, {Kuntschner}, {Davies},
  {Arimoto}, {Nakamura}, \& {Peletier}}]{vaku}
{Vazdekis} A., {Kuntschner} H., {Davies} R.~L., {Arimoto} N., {Nakamura} O.,
  {Peletier} R., 2001, \apjl, 551, L127

\bibitem[{{White}(1980)}]{whi1980}
{White} S.~D.~M., 1980, \mnras, 191, 1P

\bibitem[{{Worthey} {et~al.}(1994){Worthey}, {Faber}, {Gonzalez}, \&
  {Burstein}}]{wor1994}
{Worthey} G., {Faber} S.~M., {Gonzalez} J.~J., {Burstein} D., 1994, \apjs, 94,
  687

\bibitem[{{Worthey} \& {Ottaviani}(1997)}]{wo1997}
{Worthey} G., {Ottaviani} D.~L., 1997, \apjs, 111, 377

\end{thebibliography}

    \end{document}